\documentclass[aps,prd,fleqn,superscriptaddress]{revtex4}
\usepackage{graphicx,xcolor,natbib,braket}
\usepackage{amsmath,amssymb,amsfonts}

\newcommand{\bse}{\begin{subequations}}
\newcommand{\ese}{\end{subequations}}
\newcommand{\be}{\begin{equation}}
\newcommand{\ee}{\end{equation}}
\newcommand{\bea}{\begin{eqnarray}}
\newcommand{\eea}{\end{eqnarray}}
\newcommand{\ba}{\begin{array}}
\newcommand{\ea}{\end{array}}

\input amssym.def
\input amssym.tex

\usepackage[colorlinks=true, linkcolor=blue, bookmarks=true]
{hyperref}

\begin{document}
\title{Analytic Holographic Pseudo-Entropy in a Boosted BTZ Geometry}
\author{ M. M. Daryaei Goki\footnote{$\rm{m}_{-}$daryaeigoki@sbu.ac.ir}}
\author{M. Ali-Akbari\footnote{$\rm{m}_{-}$aliakbari@sbu.ac.ir}}
\affiliation{Department of Physics, Shahid Beheshti University, 1983969411, Tehran, Iran}

\begin{abstract}
We investigate holographic entanglement entropy and its generalization to pseudo-entropy in a boosted BTZ black hole geometry, which is holographically dual to a moving strongly coupled plasma. By solving the extremal surface equation, we derive an {\it exact closed-form expression} for the holographic entanglement entropy as a function of the boundary interval length, boost velocity, temperature, and conserved quantities. The resulting exact analytic formula correctly reproduces the standard thermal BTZ entanglement entropy in the vanishing-boost limit and smoothly reduces to the vacuum conformal result at zero temperature, thereby providing robust consistency checks. Motivated by the off-diagonal structure of the boosted metric arising from the Lorentz transformation, we then study two families of analytic continuations to Euclidean signature. Both require a simultaneous continuation of the time coordinate and the boost parameter so that the Euclidean metric remains real. If only these two parameters are continued, the boundary interval becomes imaginary and the analytically continued holographic entanglement entropy is complex-valued; we interpret this complex entropy as the holographic dual of pseudo-entropy for the boosted thermal state. If, in addition, a conserved charge is also continued, the boundary interval remains real and the resulting entropy is real. Our construction establishes a direct connection between boosted HEE and holographic pseudo-entropy, extending previous analytic frameworks developed for rotating BTZ geometries to Lorentz-boosted spacetimes.
\end{abstract}
\maketitle
\tableofcontents

\section{Introduction}
Gauge-gravity duality \cite{Maldacena:1997re,Witten:1998qj,Nastase:2007kj}, as a generalization of the AdS/CFT correspondence, relates a strongly coupled field theory to a classical gravitational system. Since standard perturbative methods are inapplicable to strongly coupled regimes, this duality serves as an indispensable tool, offering a tractable geometric description of otherwise intractable dynamics. The duality assigns geometric counterparts, often remarkably simple, to physical quantities and processes in the field theory. Notable applications include the study of thermalization and horizon formation \cite{Balasubramanian:2010ce}, the behavior of the quark--gluon plasma \cite{Casalderrey-Solana:2011dxg}, information-theoretic observables such as entanglement entropy \cite{Nishioka:2009un}  and the fundamental question of spacetime emergence \cite{VanRaamsdonk:2010pw}. Collectively, these investigations showcase the remarkable power of holographic methods, providing non-perturbative insights that bridge high-energy physics, condensed matter systems and quantum information theory.

One of the most fundamental measures in quantum information theory is entanglement entropy \cite{Nielsen:2012yss}. This quantity quantifies the degree of quantum correlations between two subsystems, effectively measuring the information loss incurred when one part of a composite system is traced out. In essence, it tells us how much we fail to know about the full system when we only have access to part of it. Formally, consider a pure state $|\psi\rangle$ partitioned into a subsystem $A$ and its complementary $\bar{A}$. The entanglement entropy is then defined as the von Neumann entropy of the reduced density matrix,
\begin{equation}
S_A = -\mathrm{Tr}\left(\rho_A \ln \rho_A\right),
\end{equation}
where $\rho_A = \mathrm{Tr}_{\bar{A}} \, |\psi\rangle\langle\psi|$. Put simply, if $A$ and $\bar{A}$ share no quantum correlations, then $S_A = 0$, meaning we lose no information by ignoring the rest. In contrast, the more entangled the two parts are, the larger this entropy becomes. For a globally pure state, the symmetry $S_A = S_{\bar{A}}$ naturally holds, reflecting the bipartite nature of the entanglement. Its value ranges from zero for separable product states to the logarithm of the Hilbert space dimension for maximally entangled states. 

Another interesting quantity that generalizes entanglement entropy and has recently attracted significant attention is pseudo-entropy \cite{Nakata:2020luh,Caputa:2024gve,Mollabashi:2021xsd,Mollabashi:2020yie}. It extends the notion of measuring correlations from a single quantum state to a pair of quantum states. Consider two arbitrary pure states $|\psi\rangle$ and $|\phi\rangle$ in the full Hilbert space, with the partition into subsystem $A$ and its complement $\bar{A}$ defined as before. The transition matrix between these two states is then defined as
\begin{equation}\label{pure}
\mathcal{T}=\frac{|\psi\rangle\langle\phi|}{\langle\phi|\psi\rangle},
\end{equation}
which is generally non-Hermitian. The reduced transition matrix for region $A$ is 
$\mathcal{T}_A=\mathrm{Tr}_{\bar{A}}(\mathcal{T})$, and the pseudo-entropy is given by 
\begin{equation}\label{pseudo}
S_{pA}=-\mathrm{Tr}_{A}\left(\mathcal{T}_A \log \mathcal{T}_A\right).
\end{equation} 
Since $\mathcal{T}$ is non-Hermitian, its eigenvalues can be complex, and consequently the pseudo-entropy may take complex values, a feature absent in standard entropic quantities. It is worth noting that the definition requires a non-zero overlap $\langle\phi|\psi\rangle$; in the limit where the two states are orthogonal, the transition matrix becomes ill-defined, signaling a singular regime for this measure. When $|\psi\rangle = |\phi\rangle$, pseudo-entropy clearly reduces to the usual entanglement entropy. Intuitively, pseudo-entropy quantifies the distinguishability or interplay between the two states $|\psi\rangle$ and $|\phi\rangle$, as effectively observed from the reduced perspective of subsystem $A$. In simple terms, while entanglement entropy tells us how region $A$ is correlated with its complement within a single state, pseudo-entropy tells us how the difference between two states is reflected within that same region.

Within the framework of gauge-gravity duality, Ryu and Takayanagi (RT) proposed a geometric prescription for computing the entanglement entropy of a spacelike subregion in a static gravitational background \cite{Ryu:2006bv}. According to this prescription, the holographic entanglement entropy (HEE) is determined by the area of an extremal surface in the bulk whose boundary coincides with the boundary of the corresponding subregion
\begin{equation}\label{RT}
S_A=\frac{\mathrm{Area}(\Gamma_A)}{4G_N},
\end{equation}
where $\Gamma_A$ denotes the extremal surface and $G_N$ is Newton's constant. In this work, we consider a (2+1)-dimensional bulk spacetime corresponding to a (1+1)-dimensional boundary field theory. Geometrically, the extremal surface is selected by minimizing the area functional among all bulk surfaces sharing the same boundary anchor, ensuring that the entropy scales with the area rather than the volume, a hallmark of holographic encoding. For time-dependent backgrounds, the RT prescription is generalized to the covariant Hubeny--Rangamani--Takayanagi formula where the extremal condition is extended to account for dynamical bulk geometries \cite{Hubeny:2007xt}. The study of HEE has since been extended to a wide range of gravitational systems \cite{Rangamani:2016dms,MohammadiMozaffar:2016vpf} and quantum-information settings \cite{Ali-Akbari:2021zsm}, including finite temperature configurations \cite{Ebrahim:2023ush}, time-dependent spacetimes \cite{Blanco:2013joa,Sahraei:2021wqn}, and boosted configurations \cite{Liu:2006nn,Mishra:2016yor,Bhatta:2019eog,Goki:2024lyf}. Collectively, these developments establish HEE as a cornerstone of the holographic dictionary, offering a concrete geometric dual for a fundamentally quantum-mechanical observable and enabling novel explorations of the interplay between geometry, quantum correlations and many-body physics.

To obtain pseudo-entropy within the gauge-gravity duality, as a natural generalization of HEE, it was first proposed in \cite{Doi:2023zaf} that an analytic continuation of the HEE formula suffices to reproduce the desired pseudo-entropy in (1+1)-dimensional conformal field theories. Essentially, this procedure extends the defining parameters of the entropy functional into the complex plane, allowing the geometric area to acquire an imaginary component that encodes the transition between two distinct quantum states. In simple terms, while the standard HEE probes the entanglement structure of a single boundary state, its analytically continued version captures the interference or overlap between two different boundary configurations as seen through the bulk geometry.

In a closely related development, timelike entanglement entropy and its holographic dual have been extensively studied in the literature \cite{Ali-Akbari:2026xzx}. Unlike the standard spacelike RT surface, the timelike prescription involves bulk surfaces with a signature change, effectively probing causal structures beyond the usual equal-time slices. Remarkably, the complex-valued holographic timelike entanglement entropy can be directly identified with pseudo-entropy. This connection is particularly elegant because it reveals that the non-Hermitian nature of the transition matrix, which gives rise to complex entropic values, finds a natural geometric counterpart in the timelike signatures of the bulk extremal surface. Consequently, these frameworks mutually enrich one another, offering complementary perspectives on how information-theoretic measures extend beyond conventional entropies into the domain of state mixing, non-equilibrium dynamics, and the deeper structure of the holographic dictionary.
 
In this work, we study HEE in a $(1+1)$-dimensional boosted strongly coupled plasma. Starting from the static BTZ black hole geometry, we construct the gravitational dual of a uniformly moving thermal state by applying a Lorentz boost along the boundary spatial direction. This boost effectively places the plasma in a moving frame, introducing a non-zero momentum density and breaking the manifest rotational symmetry of the static background. Employing the RT prescription, we derive an exact analytic expression for the HEE as a function of the arbitrary boost velocity and temperature. We subsequently examine several consistency limits, including the static case ($v=0$) and the zero temperature limit ($T=0$). In these limiting regimes, we explicitly demonstrate that our general formula reduces to the known thermal and vacuum entanglement entropies, thereby providing a robust validation of our result. Finally, we perform an appropriate analytic continuation of the HEE into the complex plane and investigate the resulting complex-valued entropy. In simple terms, this continuation amounts to extending the geometric area functional beyond real parameters, allowing the entropy to acquire an imaginary part that encodes the overlap between distinct boundary configurations. This step is particularly significant as it connects our holographic framework to the notion of pseudo-entropy, offering a concrete geometric realization of how complex entropic measures emerge from the interplay between thermal motion and quantum correlations in the dual field theory.

\section{The Static HEE}
We aim to compute the HEE for a boosted strongly coupled plasma in $(1+1)$-dimensional spacetime. Before proceeding with this calculation, however, we study simpler cases, namely the strongly coupled plasma at zero temperature and an arbitrary finite temperature $T$. This allows us to establish the basic setup and provide a clear understanding of the holographic calculations before considering the boosted configuration.
Therefore, we begin by considering a three-dimensional Anti-de Sitter ($\mathrm{AdS}_3$) black hole background, known as a BTZ black hole. The metric of the BTZ black hole in Poincar\'e coordinates is given by
\begin{equation}\label{1}
ds^2=\frac{1}{z^2}
\left(
-f(z)dt^2+\frac{dz^2}{f(z)}+dx^2
\right),
\end{equation}
where $z$ is the holographic radial coordinate. The asymptotic boundary of the spacetime is located at $z=0$, while the black hole horizon is positioned at $z=z_h$. The coordinates $x$ and $t$ correspond to the spatial and temporal coordinates of the dual field theory, respectively. The function $f(z)$ is the blackening factor of the black hole, which is given by
\begin{equation}\label{2}
f(z)=1-\frac{z^2}{z_h^2}.
\end{equation}
It is clear that the blackening factor vanishes at the horizon, $f(z_h)=0$. The Hawking temperature of the BTZ black hole, which is identified with the temperature of the dual field theory, is given by
\begin{equation}\label{3}
T=\frac{1}{2\pi z_h}.
\end{equation}
To compute the HEE, we consider the entangling region $A$ as a finite spatial interval in the boundary field theory, defined by
\begin{equation}\label{33}
A:\left\{x\in\left[-\frac{\Delta x}{2},\,\frac{\Delta x}{2}\right],\ t=0\right\},
\end{equation}
Here, $\Delta x$ denotes the size of the entangling region. Due to the symmetry of the interval around $x=0$, the corresponding bulk extremal curve reaches its maximum radial depth at $x=0$, where the turning point is located at $z=z_*$. Because the boundary theory is $(1+1)$-dimensional, the entangling region is a one-dimensional interval. According to the RT prescription, the corresponding extremal curve in the bulk is therefore a one-dimensional curve, which is a geodesic in the bulk geometry. The profile of this geodesic can be described by the embedding function $x(z)$.
The HEE is then obtained by extremizing the length functional of the geodesic
\begin{equation}\label{4}
S_{\Delta x}=\frac{L_{\Delta x}}{4G_N},
\end{equation}
where $L_{\Delta x}$ represents the length of the extremal curve.
We now need to extremize the corresponding length functional of the bulk geodesic. Using the black hole metric, the length functional can be written as
\begin{equation}\label{5}
L_{\Delta x}=\int {\cal L}_{\Delta x}\ dz=\int \frac{1}{z}
\sqrt{\frac{1}{f(z)}+\dot x(z)^2}\,dz.
\end{equation}
and  the dot denotes differentiation with respect to the radial coordinate $z$. Since $x(z)$ does not appear explicitly in the above Lagrangian, there exists a conserved quantity associated with the translational symmetry along the $x$-direction. This conserved quantity is obtained from
\begin{equation}\label{6}
P=\frac{\partial {\cal L}_{\Delta x}}{\partial \dot x}.
\end{equation}
Solving this equation for $\dot{x}(z)$, one obtains
\begin{equation}\label{7}
\dot x(z)=
\frac{z P}
{\sqrt{f(z)(1-P^2z^2)}} ,
\end{equation}
where the integration constant has been fixed by requiring that the geodesic reaches its turning point at $z=z_*$. At the turning point, the derivative $\dot x(z)$ diverges which leads to the condition
\begin{equation}\label{8}
Pz_*=1.
\end{equation}
The endpoints of the bulk geodesic are anchored on the asymptotic boundary at
\begin{equation}\label{9}
x(\epsilon)=\pm\frac{\Delta x}{2},
\end{equation}
where $\epsilon$ is the UV cutoff regulating the asymptotic boundary. Therefore, the size of the entangling region is determined by
\begin{equation}\label{10}
\frac{\Delta x}{2}=
\int_{\epsilon}^{z_*} \dot x(z)\,dz.
\end{equation}
This relation determines the connection between the turning point $z_*$ of the extremal curve and the boundary interval size $\Delta x$.
Having obtained the profile of the extremal curve, the remaining task is to evaluate its regularized length. Substituting the solution $x(z)$ into the length functional, the HEE is obtained from \eqref{4}.
In the following, we evaluate the above integrals analytically and recover the standard results for the HEE of a finite interval in a zero temperature and thermal conformal field theory. This result provides a consistency check of our setup before extending the analysis to the boosted strongly coupled plasma.

\subsection{Zero Temperature Case}
In this subsection, we first consider the zero temperature limit, corresponding to $f(z)=1$, or equivalently $z_h\rightarrow\infty$. In this limit, the black hole geometry reduces to pure $\mathrm{AdS}_3$ and the HEE can be obtained analytically. Thus \eqref{7} and \eqref{8} lead to
\begin{equation}\label{12}
\dot x(z)=\frac{z}{\sqrt{z_*^2-z^2}} .
\end{equation}
Using the boundary condition \eqref{10},
we obtain
\begin{equation}\label{14}
\frac{\Delta x}{2}=\sqrt{z_*^2-\epsilon^2}
\end{equation}
which immediately gives
\begin{equation}\label{15}
z_*=\sqrt{(\frac{\Delta x}{2})^2+\epsilon^2}.
\end{equation}
This relation shows that the maximum depth reached by the extremal curve is directly determined by the size of the entangling region. We now evaluate the regularized length of the geodesic by substituting the solution into the length functional
\begin{align}\label{15}
L_{\Delta x}&=2\int_{\epsilon}^{z_*}
\frac{dz}{z\sqrt{1-\frac{z^2}{z_*^2}}},\\ \nonumber
&=2\ln\left(\frac{\Delta x}{\epsilon}\right)+\mathcal{O}(\epsilon).
\end{align}
Finally, using \eqref{4}, and the relation between the central charge and Newton constant,
we obtain
\begin{equation}\label{19}
S_{\Delta x}=
\frac{c}{3}
\ln\left(\frac{\Delta x}{\epsilon}\right)+\mathcal{O}(\epsilon),
\end{equation}
where
$c=\frac{3}{2G_N}$.
This is the standard vacuum HEE of an interval of length $\Delta x$ in a (1+1)-dimensional conformal field theory.

\subsection{Finite Temperature Case}
We now extend the above analysis to the finite temperature case described by the BTZ black hole geometry. In this case, the blackening factor is given by \eqref{2}
and the conserved quantity obtained from the length functional leads to
\begin{equation}\label{21}
\dot x(z)=\frac{z}{\sqrt{f(z)(z_*^2-z^2)}}.
\end{equation}
Using the boundary condition,
\begin{equation}\label{23}
\frac{\Delta x}{2}=z_h\int_{\epsilon}^{z_*}\dfrac{z dz}{\sqrt{(z_h^2-z^2)(z_*^2-z^2)}}=z_h\tanh^{-1}(\dfrac{z_*}{z_h}),
\end{equation}
one obtains the relation between the turning point $z_*$ and the size of the boundary interval
\begin{equation}\label{24}
z_*
=z_h \tanh\left(\frac{\Delta x}{2z_h}\right).
\end{equation}
Substituting this result into the regularized length functional \eqref{5} gives
\begin{equation}\label{25}
L_{\Delta x}=2\ln\left[\frac{2z_h}{\epsilon}\sinh\left(\frac{\Delta x}{2z_h}\right)\right]+\mathcal{O}(\epsilon),
\end{equation}
and therefore, the HEE becomes
\begin{equation}\label{26}
S_{\Delta x}=\frac{c}{3}\ln\left[\frac{2z_h}{\epsilon}\sinh\left(\frac{\Delta x}{2z_h}\right)\right]+\mathcal{O}(\epsilon).
\end{equation}
This is precisely the well-known finite temperature HEE of an interval in a thermal (1+1)-dimensional conformal field theory. In the limit $z_h\rightarrow\infty$, or equivalently $T\rightarrow0$, this expression smoothly reduces to the vacuum result obtained in the previous subsection.

\section{Analytic HEE in a Boosted BTZ Black Hole}
In this section, we present an analytic construction of the HEE in a boosted BTZ geometry. 
To describe the dual gravitational background of a $(1+1)$-dimensional moving plasma, we perform a Lorentz boost in the $(t,x)$ plane according to
\begin{align}\label{29}
t &\rightarrow \gamma (t-vx), \\
x &\rightarrow \gamma (x-vt),
\end{align}
where $\gamma^{-1}=\sqrt{1-v^{2}}$
and $v$ denotes the velocity of the plasma. This transformation maps the static BTZ black hole to the gravitational dual of a uniformly moving thermal state in the boundary field theory.
The resulting boosted metric takes the form
\begin{equation}\label{31}
ds^{2}
=\frac{1}{z^{2}}
\left(
\frac{dz^{2}}{f(z)}
+g_{tt}\,dt^{2}
+2g_{tx}\,dt\,dx
+g_{xx}\,dx^{2}
\right),
\end{equation}
where
\begin{align}\label{32}
g_{tt}&=-\gamma^{2}(f-v^{2}),\\\label{32a}
g_{tx}&=\gamma^{2}v(f-1),\\\label{32b}
g_{xx}&=\gamma^{2}(1-fv^{2}).
\end{align}
The Lorentz boost generates the off-diagonal metric component $g_{tx}$, indicating that the boosted geometry is no longer static in these coordinates. As a consequence, the temporal and spatial directions become coupled, making the determination of the extremal surface considerably more involved than in the static background.
The boundary subsystem whose HEE we wish to compute is described by \eqref{33}. Because of the non-vanishing off-diagonal component of the metric, the extremal surface cannot be described solely by the embedding function $x(z)$. Instead, it must be parametrized by two independent embedding functions $x=x(z)$ and $t=t(z)$ which are determined simultaneously by extremizing the area functional. Consequently, the variational problem becomes more complicated than in the static case.
The corresponding length functional is
\begin{equation}\label{36}
{\cal L}_{\Delta x}
=
\frac{\sqrt{Q}}{z},
\end{equation}
where
\begin{equation}\label{37}
Q
=
\frac1f
+\gamma^{2}(v^{2}-f)\dot t^{\,2}
+2\gamma^{2}v(f-1)\dot t\,\dot x
+\gamma^{2}(1-fv^{2})\dot x^{\,2}.
\end{equation}
Since neither $t(z)$ nor $x(z)$ appears explicitly in the Lagrangian, the translational symmetries generated by the Killing vectors $\partial_t$ and $\partial_x$ imply the existence of two conserved quantities
\begin{align}\label{38}
E&=\frac{\partial\mathcal{L}_{\Delta x}}{\partial\dot t}
=\frac{\gamma^{2}}{z\sqrt Q}\left[(v^{2}-f)\dot t+v(f-1)\dot x\right],\\
P&=\frac{\partial\mathcal{L}_{\Delta x}}{\partial\dot x}=\frac{\gamma^{2}}{z\sqrt Q}\left[(1-fv^{2})\dot x+v(f-1)\dot t\right].\label{38a}
\end{align}
These conserved quantities considerably simplify the analysis. In the following, we solve these equations analytically, determine the turning point of the extremal surface and obtain a closed-form expression for the HEE in the boosted background.
Solving \eqref{38} and \eqref{38a} for the embedding derivatives, we obtain
\begin{align}\label{39}
\dot t
&=
\frac{\gamma^{2}z\sqrt{Q}}{f}
\Big[
(1-fv^{2})E
-
v(f-1)P
\Big],
\\
\dot x
&=
\frac{\gamma^{2}z\sqrt{Q}}{f}
\Big[
-v(f-1)E
-
(v^{2}-f)P
\Big]\label{39a}.
\end{align}
Although these expressions determine the embedding functions, they remain coupled through the quantity $Q$, which still depends explicitly on both $\dot t$ and $\dot x$. To eliminate this dependence, we substitute the above expressions into \eqref{37}. After straightforward algebra, one finds
\begin{equation}\label{40}
Q
=
\frac{1}{f
+
\gamma^{2}z^{2}s(z)},
\end{equation}
where
\begin{equation}\label{41}
s(z)
=
(1-fv^{2})E^{2}
+
(v^{2}-f)P^{2}
-
2v(f-1)EP.
\end{equation}
This result expresses $Q$ entirely in terms of the conserved quantities. The turning point of the extremal surface is determined by the condition that the derivatives $\dot t$ and $\dot x$ diverge simultaneously. This occurs when the denominator of the above expressions vanishes, yielding
\begin{equation}\label{42}
f(z_*)+\gamma^{2}z_*^{2}s(z_*)=0.
\end{equation}
Solving this equation gives
\begin{equation}\label{433}
z_*^{2}=z_h^2\frac{-\left(E^{2}-P^{2}-\dfrac1{z_h^{2}}\right)\pm
\sqrt{\left(E^{2}-P^{2}-\dfrac1{z_h^{2}}\right)^{2}-
\dfrac{4\gamma^{2}}{z_h^{2}}(P+Ev)^{2}}}{
2\gamma^{2}(P+Ev)^{2}}.
\end{equation}
Since the turning point must remain real for a spacelike extremal surface, two conditions must be satisfied simultaneously:
\begin{itemize}
\item The following condition must hold:
\begin{equation}\label{45}
E^{2}-P^{2}-\frac{1}{z_{h}^{2}}<0.
\end{equation}
The turning point equation \eqref{433} then admits two algebraic branches. In the static limit $v\rightarrow 0$ and hence $E\rightarrow 0$, \eqref{433} must reduce to the familiar turning-point equation of the static BTZ geometry. Evaluating this limit, one finds that the branch corresponding to the negative sign smoothly reduces to the physical turning point \eqref{8}. 
In contrast, for the branch with the positive sign, the turning point approaches the horizon and therefore does not represent a physical turning point. Instead, it appears as an auxiliary root entering the factorization of the quartic polynomial in the radial integrals. Throughout this work, we therefore identify the {\it{negative branch}} as the physical solution.

\item In order to have a real root, the quantity under the square root in \eqref{433} must be non-negative
\begin{equation}\label{43}
\left(E^{2}-P^{2}-\frac{1}{z_{h}^{2}}\right)^{2}-
\frac{4\gamma^{2}}{z_{h}^{2}}(P+Ev)^{2}\ge 0.
\end{equation}
\end{itemize} 
Together, these conditions ensure that the turning point is real and that the corresponding extremal surface is physically admissible. This is precisely the condition $z_*\le z_h$ obtained for the spacelike geodesic in the static BTZ geometry.

Having determined the turning point, we next impose the boundary conditions associated with the spacelike interval. Substituting \eqref{40} into \eqref{39} and  \eqref{39a}, and integrating from the turning point to the asymptotic boundary, we obtain
\begin{align}\label{47}
0=\frac{\Delta t}{2\gamma^2}
&=
P\int_{\epsilon}^{z_*}\left[\dfrac{v(f-1)z}{f\sqrt{D}}\right]dz
-
E\int_{\epsilon}^{z_*}\left[\dfrac{(1-fv^2)z}{f\sqrt{D}}\right]dz,
\\
\frac{\Delta x}{2\gamma^2}
&=
E\int_{\epsilon}^{z_*}\left[\dfrac{v(f-1)z}{f\sqrt{D}}\right]dz
-
P\int_{\epsilon}^{z_*}\left[\dfrac{(v^2-f)z}{f\sqrt{D}}\right]dz,
\label{47a}
\end{align}
where $D=f+(E^{2}-P^{2})z^{2}+\frac{\gamma^{2}}{z_h^{2}}
(P+Ev)^{2}z^{4}$.
The boundary conditions given by \eqref{47} and \eqref{47a} are not independent. The first equation can be used to eliminate the remaining integral appearing in the second one. As a result, the boundary interval is expressed in terms of a single radial integral,
\begin{equation}\label{49}
\frac{\Delta x}{2}
=
\int_{\varepsilon}^{z_*}
\frac{z}{f\sqrt{D}}
\left[
\left(
\frac{E^{2}}{P}+P
\right)
-
\left(
P-\frac{E^{2}v^{2}}{P}
\right)
\frac{z^{2}}{z_h^{2}}
\right]dz .
\end{equation}
This representation considerably simplifies the problem, since only two elementary integrals remain to be evaluated. Performing the integration, we obtain
\begin{equation}\label{50}
\frac{\Delta x}{2}
=
\frac{z_h}{P\gamma}
\left[
(P-Ev)\tanh^{-1}\!\left(\frac{z_*}{z_r}\right)
-
(E-Pv)\tanh^{-1}(X)
\right],
\end{equation}
where
\begin{equation}\label{51}
X=
\frac{\sqrt{z_h^{2}-z_r^{2}}}
{\sqrt{z_h^{2}-z_*^{2}}}
\frac{z_*}{z_r},
\end{equation}
and $z_r$ denotes the auxiliary real root of \eqref{433}.
The above relation can be inverted straightforwardly as
\begin{equation}\label{52}
\frac{z_*}{\sqrt{z_r^{2}-z_*^{2}}}
=
\sinh\!\left[
\frac{P\gamma}{2z_h}
\frac{\Delta x}{P-Ev}
+
\frac{E-Pv}{P-Ev}
\tanh^{-1}(X)
\right].
\end{equation}
Having established the relation between the turning point and the boundary interval, we now evaluate the regularized length of the extremal surface,
\begin{equation}\label{53}
L_{\Delta x}
=
\int_{\epsilon}^{z_*}
\mathcal{L}_A\,dz
=
\ln\!\left(
\frac{2z_r}{\epsilon}
\frac{z_*}{\sqrt{z_r^{2}-z_*^{2}}}
\right)+{\cal O}(\epsilon).
\end{equation}
Substituting \eqref{52} into the above expression, we obtain
\begin{equation}\label{54}
L_{\Delta x}
=
\ln\!\left[
\frac{2z_r}{\epsilon}
\sinh\!\left(
\frac{P\gamma}{2z_h}
\frac{\Delta x}{P-Ev}
+
\frac{E-Pv}{P-Ev}
\tanh^{-1}(X)
\right)
\right]+{\cal O}(\epsilon).
\end{equation}
According to the RT prescription, the HEE is related to the regularized geodesic length through
$S_{\Delta x}=\frac{c}{3}L_{\Delta x}$
which immediately gives
\begin{equation}\label{56}
S_{\Delta x}=
\frac{c}{3}
\ln\!\left[
\frac{2z_r}{\epsilon}
\sinh\!\left(
\frac{P\gamma}{2z_h}
\frac{\Delta x}{P-Ev}
+
\frac{E-Pv}{P-Ev}
\tanh^{-1}(X)
\right)
\right]+{\cal O}(\epsilon).
\end{equation}
The analytic expression \eqref{56} admits two nontrivial consistency checks as follows:
\begin{itemize}
\item We consider the thermal limit, meaning that $v\rightarrow 0$, $E\rightarrow 0$ and $z_r\rightarrow z_h$.
Substituting these limits into \eqref{56} yields \eqref{26}.

\item We consider the zero temperature limit, meaning that $z_h\rightarrow \infty$ and $E\rightarrow 0$.
Using \eqref{43} and \eqref{51}, one finds $X\rightarrow 0$ and $z_r\rightarrow \dfrac{z_h}{\gamma}$ so that \eqref{56} becomes
the vacuum result \eqref{19}. 
\end{itemize}

The HEE for a Lorentz-boosted subsystem has already been investigated in \cite{Kusuki:2017jxh}. In particular, the case of a boosted interval, related to a constant-time slice and a light-like interval in AdS$_{d+1}$, dual to a CFT$_{d}$, has been studied. There, the authors start from a (thermal) AdS$_{d+1}$ background and apply a boost to the interval. In our case, however, we consider the space-like interval \eqref{33} and compute the HEE in a boosted AdS$_3$ geometry.
Although our computation and theirs are closely related, and although one expects, by appropriate Lorentz transformation and the   identification of boundary data, that the static BTZ geometry with a boosted interval should yield the same HEE as the boosted BTZ geometry with a static interval, there is unfortunately no analytical result in that work that can be directly compared with our final expression \eqref{56}. Indeed, no analytic solution is presented there; instead, for a given temperature, the HEE is plotted as a function of $\Delta x$ and $\Delta t$, with $v=\Delta x/\Delta t$. In our approach, by contrast, the final result is expressed in terms of $\Delta x$, $v$, $E$, and $P$, which must satisfy the constraints \eqref{45}, \eqref{43} and \eqref{50}. This allows us to obtain a closed-form expression and to exhibit the full family of extremal surfaces, which is not accessible from the purely numerical treatment of that work.

\begin{figure}[htbp]
\includegraphics[width=88 mm]{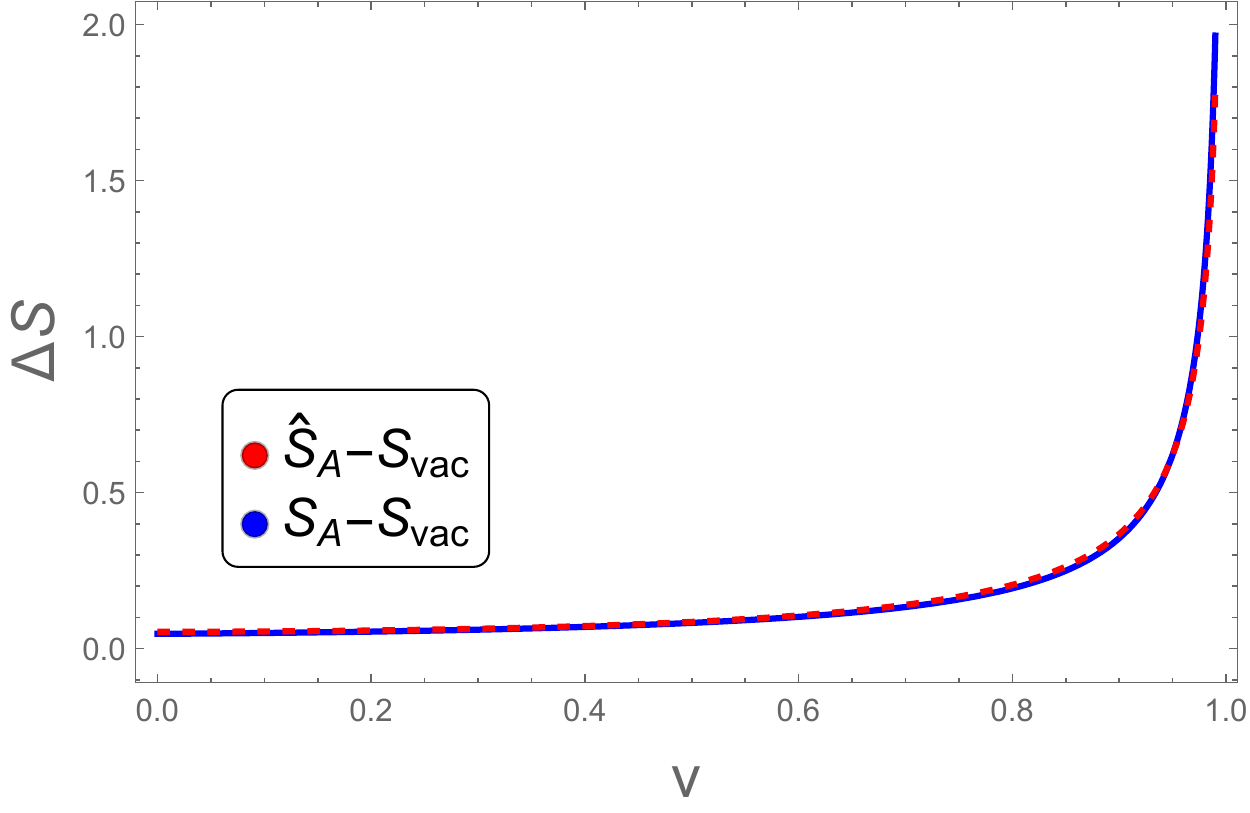}  
\caption{Comparison of the HEE obtained from \eqref{1000} (the red dashed curve) and \eqref{56} (the blue curve) for $P=100$, $E=20$ $z_h=1$, and $\Delta x=2$. Here $S_{\rm vac}$ is the HEE at zero temperature and zero velocity.}
\label{fig1b}
\end{figure}

For the boosted BTZ black hole, an analytic expression for the HEE was previously derived in \cite{Bhattacharya:2022msw} by applying a Lorentz boost to the static thermal entropy solution. Their result takes the form
\begin{equation}\label{1000}
\hat{S}_{\Delta x}=\frac{c}{6}\ln\left[\frac{\beta^2}{\pi^2 \epsilon^2}
\sinh\left(\dfrac{\pi \alpha \Delta x}{\beta}\right)\sinh\left(\dfrac{\pi \Delta x}{\alpha\beta}\right)\right],
\end{equation}
where the boost parameter is encoded in $\alpha^2(1+v)=1-v$ and $\beta=T^{-1}$. However, our derivation is formulated directly in terms of the conserved quantities associated with the extremal surface. Consequently, the final analytic expression \eqref{56} is written explicitly in terms of $\Delta x$, $v$ and the constants $E$ and $P$ for given temperature. Thus, in order to compare our result with theirs, one would need to establish a relation between the pair $(E,P)$ and the boundary data $(\Delta x,v)$, which in our case is given by the rather complicated relations \eqref{47} and \eqref{47a}. Although it seems difficult to find an analytical relation between these two pairs, interestingly the HEE \eqref{1000} is reproduced, at least numerically, in the limit $P\gg E$ of \eqref{56}. Therefore, this suggests that these two HEEs describe the same physics but in terms of different parameters, with the relation between them remaining unclear. Figure~\ref{fig1b} presents a representative comparison between the two analytic formulas, demonstrating their excellent agreement.

In short, although the same configuration has been investigated in \cite{Bhattacharya:2022msw} and \cite{Kusuki:2017jxh}, there are two important points that make our work valuable. First, the calculation is carried out here step by step and in a clear manner, yielding an {\it analytical result} expressed in terms of the underlying parameters. Second, our final result is formulated in terms of the constants of motion $E$ and $P$, which are the conserved momenta. This formulation offers two advantages: it makes the dependence of the HEE on the conserved charges explicit, and it provides a convenient handle for comparing with the known result of \cite{Bhattacharya:2022msw}, which is recovered in the limit $P\gg E$. Moreover, expressing the result in terms of $E$ and $P$ makes the analytic continuation to Euclidean signature, and hence the construction of the holographic pseudo-entropy, particularly transparent, since the continuation acts directly on the conserved quantities.

Having established that the boosted HEE correctly reproduces all physically expected limits, we now turn to its analytic continuation and the corresponding holographic pseudo-entropy.

\section{Holographic Pseudo-entropy}
Having obtained the exact analytic expression for the HEE in the boosted BTZ black hole, we now turn to its analytic continuation. As discussed in the previous section, the boosted BTZ metric, given in (\ref{31})--(\ref{32b}), contains a non-vanishing off-diagonal component generated by the Lorentz boost. Consequently, a naive analytic continuation of the metric presents an immediate subtlety. If one simply replaces the time coordinate as \(t \to i t_E\), the off-diagonal component \(g_{tx}\) acquires an imaginary factor, rendering the metric complex-valued. Since the Euclidean metric must remain real in order to describe a well-defined Euclidean geometry, we must also analytically continue another parameter appearing in the metric, namely the boost velocity \(v \to i v_E\). It is then straightforward to verify that the resulting metric is indeed real, yielding
\begin{equation}
ds_E^{2}=\frac{1}{z^{2}}\left[\frac{dz^{2}}{f}+\gamma_E^{2}(f+v_E^{2})dt_E^{2}
-2\gamma_E^{2}v_E(f-1)dt_E\,dx+\gamma_E^{2}(1+fv_E^{2})dx^{2}\right],
\end{equation}
where \(\gamma_E^{-1}=\sqrt{1+v_E^{\,2}}\).
Crucially, upon this continuation, the signs of the off-diagonal component and of \(g_{tt}\) change relative to the original Lorentzian form. In simple terms, this sign flip effectively reverses the fictitious rotation introduced by the boost. This is a characteristic feature of the Euclidean continuation and is directly analogous to the analytic continuation employed for the rotating BTZ black hole \cite{Doi:2023zaf,Dai:2026tmp}.

We now consider two families of analytic continuation. First, we consider the case in which \(\Delta x\) remains real after the above continuations of time and velocity. In this case, we must also analytically continue the conserved quantity as \(E\rightarrow i E_E\). If, instead, we allow the interval \(\Delta x\) to become imaginary, then no analytic continuation of the constants \(E\) and \(P\) is necessary. We discuss each case separately in the following.

\begin{itemize}
\item \(t \rightarrow i t_E\), \(v \rightarrow i v_E\), and \(E\rightarrow i E_E\):

One may additionally require the boundary spatial interval \eqref{50} to remain real under the analytic continuation. While the continuations
\(t\rightarrow i t_E\) and \(v\rightarrow i v_E\) are sufficient to keep the bulk metric real, the reality of the boundary separation \(\Delta x\) requires the conserved quantity to be continued as well,
\(E\rightarrow iE_E\). This continuation leads to \(z_r^E>z_h\) and hence \(X\rightarrow iX_E\), so that
\(\tanh^{-1}(iX_E)=i\tan^{-1}(X_E)\).
The additional factor of \(i\) is then compensated by the continued conserved quantity, and the resulting expression for \(\Delta x\) remains real. The corresponding HEE takes the form
\begin{equation}
S^E_{\Delta x}
=
\frac{c}{3}
\ln\left[
\frac{2z^E_r}{\epsilon}
\sinh\left(\mathcal{S}\right)
\right],
\label{eq:SE_real}
\end{equation}
where
\begin{equation}
\mathcal{S}=
\frac{P\gamma_E}{2z_h}
\frac{\Delta x}{P+E_Ev_E}
-
\frac{E_E-Pv_E}{P+E_Ev_E}
\tan^{-1}(X_E).
\label{eq:S_real}
\end{equation}
Thus, the entropy is manifestly real,
\begin{equation}
\operatorname{Im}S_{\Delta x}^{E}=0.
\end{equation}
This shows that the analytic continuation can be implemented while keeping both the bulk geometry and the boundary spatial subregion real. This result is similar to that discussed in \cite{Doi:2023zaf}, where for a spacelike interval the imaginary part of the HEE vanishes.

\item \(t \rightarrow i t_E\) and \(v \rightarrow i v_E\):

Under these continuations, the HEE given in \eqref{56} leads to
\begin{equation}
S_{\Delta x}^E=\frac{c}{3}\ln\left[\frac{2z_r}{\epsilon}\sinh{\mathcal{S}}\right],
\end{equation}
where
\begin{equation}
\mathcal{S}=\frac{P\gamma_E}{2z_h}\frac{\Delta x}{P-iEv_E}
+\frac{E-iPv_E}{P-iEv_E}\tanh^{-1}(X).
\end{equation}
This complex-valued entropy is precisely the holographic manifestation of pseudo-entropy, where the imaginary part encodes the interplay between quantum state overlaps in the dual field theory, thus completing our geometric construction of analytically continued entropic observables.

Since the argument of the hyperbolic sine is now complex, it is convenient to decompose it into its real and imaginary parts as
\begin{equation}
\mathcal{S}=\mathcal{S}_R+i\mathcal{S}_I,
\end{equation}
with
\begin{align}
\mathcal{S}_R&=\frac{P^{2}\gamma_E}{2z_h(P^{2}+E^{2}v_E^{2})}\,\Delta x+
\frac{EP(1+v_E^{2})}{P^{2}+E^{2}v_E^{2}}\tanh^{-1}(X),\\
\mathcal{S}_I&=\frac{EPv_E\gamma_E}{2z_h(P^{2}+E^{2}v_E^{2})}\,\Delta x-
\frac{v_E(P^{2}-E^{2})}{P^{2}+E^{2}v_E^{2}}\tanh^{-1}(X).
\end{align}
Using the identity \(\sinh(x+iy)=\sinh x\cos y+i\cosh x\sin y\), the pseudo-entropy naturally separates into its real and imaginary parts, which are explicitly given by
\begin{align}
\operatorname{Re}(S_{\Delta x}^E)&=\frac{c}{3}\ln\left[\frac{2|z_r|}{\epsilon}\sqrt{\sinh^{2}(\mathcal{S}_R)\cos^{2}(\mathcal{S}_I)+
\cosh^{2}(\mathcal{S}_R)\sin^{2}(\mathcal{S}_I)}\right],\\
\operatorname{Im}(S_{\Delta x}^E)&=\frac{c}{3}\tan^{-1}\left(
\frac{\cosh(\mathcal{S}_R)\sin(\mathcal{S}_I)}{\sinh(\mathcal{S}_R)\cos(\mathcal{S}_I)}
\right).
\end{align}
It is worth noting that, as is evident from the above expressions, in the limit where the parameter \(v_E\) is set to zero, the resulting pseudo-entropy reduces precisely to \eqref{26}, which corresponds to the thermal entanglement entropy. The emergence of a non-vanishing imaginary contribution demonstrates that the analytically continued entropy is intrinsically complex. Following the interpretation proposed for the rotating BTZ black hole~\cite{Doi:2023zaf}, we identify this complex quantity as the holographic pseudo-entropy associated with the boosted BTZ geometry.

Our construction therefore provides an analytic realization of holographic pseudo-entropy in a Lorentz-boosted background. Starting from the exact analytic expression for the boosted HEE, we have shown that an appropriate Euclidean analytic continuation naturally generates a complex entropy whose real and imaginary parts admit closed analytic forms. This establishes a direct connection between boosted HEE and holographic pseudo-entropy, extending the analytic framework previously developed for rotating BTZ geometries to Lorentz-boosted spacetimes.
\end{itemize}

\section*{Acknowledgments}
We acknowledge DeepSeek for its helpful assistance in improving the clarity and presentation of this manuscript.

\end{document}